\documentclass[aps,twocolumn]{revtex4}
\usepackage[dvips]{graphicx}
\begin{document}
\draft{}
\title{Bosonic stimulation of cold 1s excitons into a 
harmonic potential minimum in Cu$_2$O}
\author{N. Naka and N. Nagasawa}
\address{Department of Physics, Graduate School of Science, 
University of Tokyo\\
7-3-1 Hongo, Bunkyo-ku, Tokyo 113-0033, Japan}
\date{17 December, 2002}

\begin{abstract}
Density distribution of cold exciton clouds generated into a strain-induced 
potential well by two-photon excitation in Cu$_2$O is studied at 2 K.
We find that an anomalous spike, which
can be interpreted as accumulation of the excitons into the ground state,
emerges at the potential minimum.
The accumulation can be due to stimulated scattering of cold excitons, 
mediated by acoustic phonon emission. Possibility of the formation of the
thermodynamic Bose-Einstein condensate of paraexcitons has been discussed.
\end{abstract}


\maketitle

Before the first realization of atomic Bose-Einstein condensation (BEC) 
in 1995 \cite{cornell}, excitons in semiconductors were thought to
be one of the most favorable candidates for realizing BEC \cite{Book}.
It has long been believed that excitons in Cu$_2$O is the best system 
to pursue such a phase transition in a three-dimensional system,
because of their long lifetime derived from the dipole forbidden gap;
the orthoexcitons couple with photons
only via quadrupole interaction, and paraexcitons are basically inactive
for optical transitions, thereby leading to a long lifetime.

However,
the long lifetime of the excitons in Cu$_2$O allows them to expand 
in free crystals, and means that their recombination rate is low. 
The former reduces the exciton density against the BEC,
and the latter makes it difficult to monitor exciton densities 
by optical means.
In order to overcome these unfavorable situations,
we employed inhomogeneous strain that creates
potential wells and accelerates the direct recombination rate. 
We have been studying kinetics of a cold exciton system in a potential well,
by using two-photon excitation. In this regime,
orthoexcitons are created near the local momentum zero \cite{PRB1,PRB2}.
In this communication, we report on a novel phenomenon
 showing stimulated scattering of excitons; this can be 
mediated by acoustic phonon emission.

\begin{figure}
\begin{center}
\includegraphics[width=7cm,clip]{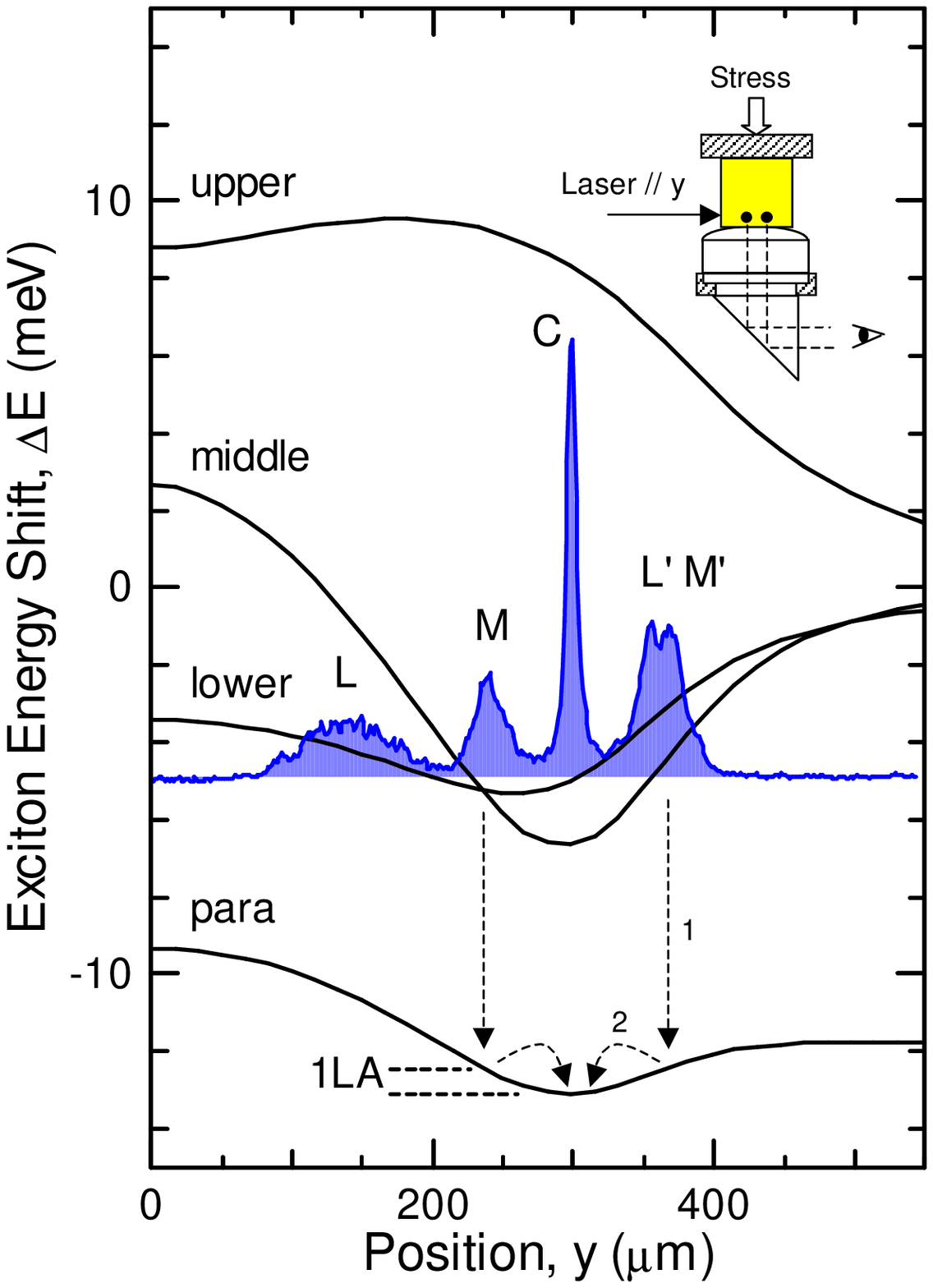}
\end{center}
\caption{Solid lines: calculated potential energies of the four 1s exciton
branches, i.e. upper-, middle-, and lower-branch orthoexcitons, and 
paraexcitons. The zero of the energy was taken to be the orthoexciton
resonance under zero stress, 2032.78 meV.
The inset shows schematic for the experimental configuration.
Two potential minima were created near the center of the contact to
a glass plunger, while data for positive $y$ are shown.
The shadowed profile shows spatial distribution of the exciton cloud,
measured at $t=10$ ns. Dashed arrows labeled 1 and 2 
show the proposed mechanism
of stimulated scattering after downconversion of orthoexcitons
to paraexcitons at the rim (see the text for detail).
}
\label{ExpSetup}
\end{figure}

A natural Cu$_2$O crystal cut into a 3 mm-cube 
was pushed against a glass lens of curvature $R=7.78$ mm 
to create harmonic potential wells (see the inset of Fig. 1).
Solid lines in Fig. 1 show calculated potential shapes
 for the four lowest exciton states in Cu$_2$O, i.e. upper-, middle- and 
lower-branch orthoexcitons and paraexcitons. 
It is evident that middle- and lower-branch orthoexcitons as well as 
the paraexcitons 
become trapped states while the upper-branch orthoexcitons form
an untrapped state.
The trapping potential shapes are approximately harmonic around the 
center of the wells.
An infrared laser beam from a color-center laser (Solar, LF151) pumped 
by a Q-switched YAG laser (SantaFe Laser, C-140)
was used to inject orthoexcitons into a potential well.
The repetition rate and the duration of the laser light were
400 Hz and 12 ns, respectively.
The two-photon excitation energy of the laser light was tuned  
slightly below the orthoexciton resonance under zero stress,
so that middle- and lower-branch orthoexcitons were generated into the trap
\footnote{
The two-photon excitation of paraexcitons is basically possible,
but is not practical to generate a high-density gas of excitons 
\cite{PRB2}.}.
The luminescence was viewed along the stress axis
through a prism mounted below the crystal.
A $\times$20 magnified image of the luminous cloud 
 was detected by an ICCD camera system (LaVision, Picostar HR12)
after being passed through a filter which cuts off the
infrared scattering light.
By inserting a spectrometer before the ICCD camera system,
we confirmed that luminescence light consists of intrinsic 
free-exciton recombination luminescence.
The spatial resolution of the optical system was higher than 10 $\mu$m.

Figure 2 shows time-resolved profile of the exciton clouds
at different times, $t$, after the incidence of the laser light.
The gate width was set at 4 ns, and respective profiles were extracted
from luminescence images averaged over 4000 laser shots.
Parts (a) and (b) show growing and decaying of the exciton density,
 with time interval of 2 ns. Three major peaks are seen in a range
 between $y=200$ $\mu$m and $y=400$ $\mu$m.
The intensity of the side peaks was almost
proportional to the temporal profile of the laser pulse, and
to the square of the power density of the laser light. On the other hand,
the central peak shows rapid growth even after 
the maximum of the incident laser intensity at $t=6$ ns,
and has nonlinear dependence on the incident power density.
The width of the central peak was 7.7 $\mu$m, much narrower than
that of side peaks.

\begin{figure}
\begin{center}
\includegraphics[width=8cm,clip]{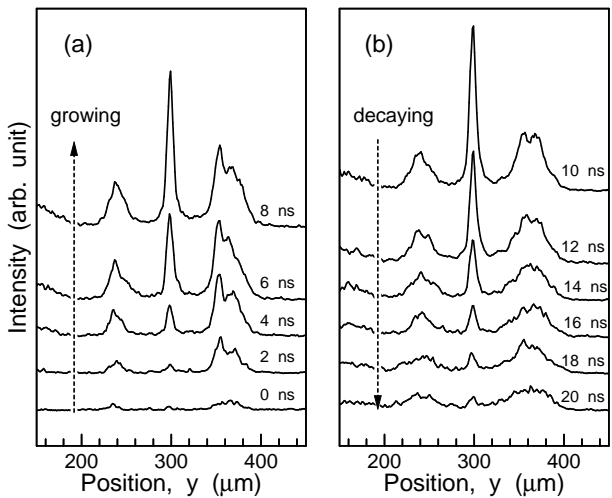}
\end{center}
\caption{
Time evolution of the spatial profile of the exciton cloud pumped
at a rim of a potential well. The baseline for each profile
 has been shifted for clarity. 
A sharp spike emerges at $y \simeq 300$ $\mu$m, and
rapidly grows after $t=4$ ns. Part (b) shows profiles after $t=10$ ns.
The gate width was set at 4 ns, and the power of the incident light
was $I=9$ mW on the sample surface.}
\label{ExpSetup}
\end{figure}

In order to understand the origin of the side peaks and of
the anomalous spike in between,
the spatial profile at $t=10$ ns is superimposed to the potential shapes 
in Fig. 1. The position of the baseline represents 
the two-photon excitation energy, measured from the orthoexciton 
resonance under zero stress, 2032.78 meV.
Resonance creation of orthoexcitons occurs in selected regions 
where the two-photon excitation energy coincides with the local exciton energy;
structures marked L, M and L', M' are due to the resonance with 
lower- and middle-branch orthoexcitons at a rim of the trap. 
On the other hand, emergence of the sharp spike, C, at the center of the well
 cannot be explained by resonance to any orthoexciton branches. 
This difference, from the case for side peaks, is consistent with the 
above features, i.e. nonlinearity and the delay in the growing process 
for the central spike, C.

The position of the center of the well for middle-branch orthoexcitons
is almost the same as that for paraexcitons. Therefore, judging from its position, the origin of the spike can be middle-branch orthoexcitons, 
paraexcitons, or both. 
Unfortunately, however, we could not obtain spectrally resolved data
in the present experiment, because of the weakness of the signal.
Dashed arrows in Fig. 1 show a possible mechanism for the 
emergence of the spike: the orthoexciton-paraexciton downconversion
occurs in the injection spot at a rim (the arrow labeled 1), 
and then paraexcitons at the rim are scattered into the center of the well
(the arrow labeled 2). 
The energy transfer necessary for this scattering process is inferred
to be 0.6 meV, in agreement with the energy position where the
acoustic (LA)-phonon dispersion crosses the exciton dispersion.
Therefore, the scattering can be mediated by acoustic phonon emission.
In fact, we found that the central spike appears only when
the laser beam traverses the center of the well, and when 
the two-photon excitation energy is set at $\Delta E=-5.03 \pm$ 0.25 meV. 
Considering the timescale of the downconversion rate, 3 ns \cite{WeinerYu},
and the energy transfer, 1.0 meV or more, necessary for the
scattering of orthoexcitons from the rim to their potential minimum, 
condensation of the middle-branch orthoexcitons is not likely.
To conclude,
the spike marked C is most likely to be due to paraexcitons 
accumulated in the ground state at the center of the potential well.

The role of the acoustic phonon in the accumulation of the condensate
has been discussed in Refs. \cite{OMSchmitt,ASchmitt}. 
The building time has been estimated as a few nanoseconds.
The observed delay is in quantitative agreement with this prediction,
provided that the central spike is due to a condensate.
Using the measured size of the central spike, the phase-space density
of paraexcitons was estimated as $\geq 1$.
A more detailed measurement of the building time of the spike 
deserves future study. Confirmation with higher spectral resolution,
 which enables measurements of the size of the ground-state
wavefunction, is in progress.

We would like to thank Mr. T. Ueda (Mrubun Co.)
and Prof. M. Kuwata-Gonokami (Univ. Tokyo)
for their kind arrangement for using the ICCD camera system,
and Mr. T. Nishi and Mr. T. Mizuno (Marubun Co.) 
for their technical support. 
This work was partially supported by The Mitsubishi foundation and
the grant-in-aid for scientific research from the Ministry of 
Education, Science and Culture, Japan.

\end{document}